# Performing Mathematical Operations using High-Index Acoustic Metamaterials

Farzad Zangeneh-Nejad and Romain Fleury*

*Laboratory of Wave Engineering, EPFL, 1015 Lausanne, Switzerland*
*To whom correspondence should be addressed. Email: romain.fleury@epfl.ch*

**The recent breakthrough in metamaterial-based optical computing devices [Science 343, 160 (2014)] has inspired a quest for similar systems in acoustics, performing mathematical operations on sound waves. So far, acoustic analog computing has been demonstrated using thin planar metamaterials, carrying out the operator of choice in Fourier domain. These so-called filtering metasurfaces, however, are always accompanied with additional Fourier transform sub-blocks, enlarging the computing system and preventing its applicability in miniaturized architectures. Here, employing a simple high-index acoustic slab waveguide, we demonstrate a highly compact and potentially integrable acoustic computing system. The system directly performs mathematical operation in spatial domain and is therefore free of any Fourier bulk lens. Such compact computing system is highly promising for various applications including high throughput image processing, ultrafast equation solving, and real time signal processing.**

## 1. Introduction

The idea of analog computation dates back to the early 19[th] century, when a number of mechanical and electronic computing machines intended for simple mathematical operations such as differentiation or integration were developed [1,2]. Such analog computing devices were then totally overshadowed by the emergence of their digital counterparts in the second half of 20[th] century, as they could not compete with the speed and reliability of digital data processors [3]. The recent breakthrough in the seemingly unrelated field of metamaterials [4-20], however, is bringing them back into the competition as an important alternative computational approach at the hardware level, with highly promising applications in ultrafast equation solving [21,22], real-time and continuous signal processing [23], and imaging [24].

In their inspiring proposal, Silva, et.al [25] proposed the first realization of computational metamaterials, performing various mathematical operations including differentiation, integration and convolution on electromagnetic waves. Such wave-based computational scheme then formed the basis for chains of ultrafast and highly efficient optical computing devices [26-39], all allowing one to go beyond the limitations of conventional analog computers.

In a general classification, all of the proposed metamaterial-based computing devices can be divided into two different categories according to the approach that is taken for the computation. In the first approach, the operator of choice is carried out by a thin planar metamaterial in Fourier domain. This approach, usually known as metasurface (MS) approach in the literature [25,26], involves the use of two additional graded index (GRIN) bulk lenses applying Fourier transform (FT) on the input and output signals of the filtering metasurface. The inevitable use of GRIN lenses drastically increases the overall size of these types of computing systems, in turn hindering their applicability in compact architectures. The second approach, usually regarded as Green's function (GF) method, however, overcomes this limitation by not doing the computation in Fourier domain. In fact, as its name suggests, the approach relies on engineering the metamaterial block itself so that its Green's function follows that of the operator of choice. Although the available bandwidth when using this approach is often substantially reduced, the resulting computing device is highly compact and potentially integrable as it avoids the usage of FT sub-blocks [25,31].

Motivated by the renewed interest in optical analog computing, there has been a quest for similar acoustic systems carrying out mathematical operations on sound waves. As a matter of fact, some acoustic computational devices have already been demonstrated utilizing transmissive [40] or reflective metasurfaces [41], offering platforms for efficient acoustic wave manipulation [42,43]. The underlying computing principle in such systems, however, relies on MS method. Consequently, the filtering metasurface is always accompanied with additional focusing metasurfaces taking the role of FT blocks, a fact which enlarges the resulting acoustic device as mentioned and, in turn, prevents its practical and flexible usage in miniaturized systems.

Here, we propose and demonstrate an acoustic computing system whose underling working principle is based on GF method, meaning that the metamaterial block carries out the computation directly in spatial domain, without any need for additional Fourier lenses. We show how the proposed system is capable of performing different mathematical operations such as first or second order differentiation, and integration. We further demonstrate the possible usage of our computing device for edge detection of an image. Compared with the reported MS-based devices having overall lengths of about $L = 10\lambda$ or more, our computing acoustic system is comparable in its length with the wavelength of operation, allowing its practical and flexible usage for the synthesis of highly efficient and compact architectures. Altogether, our results constitute a framework for various innovative acoustic applications comprising medical imaging, ultrafast equation solving, and real time signal processing.

## 2. Methods

To start, consider the geometry of Fig. 1a, depicting a metamaterial built from intercrossing air-filled acoustic pipes with the radius *R= 1.5 cm*, arranged in a square lattice with the lattice constant *a=5 cm*. Such kind of metamaterial coils up the space and provides an internal geometrical detour for sound, reducing its effective travelling velocity [44-47]. Consequently, from a macroscopic point of view, the structure effectively acts as a high-index acoustic medium, a result which was already explored in [47]. For further assertion, however, we have calculated the band structure of the crystal using finite-element simulations and represented it in Fig. 1b. Inspecting the obtained band structure reveals that the metamaterial under investigation indeed imitates a medium with the refractive index of $n_{eff}>1$, as its dispersion curve falls below the sound cone and varies linearly within the frequency range of 0 and 1 *KHz*. Now, consider a finite piece of the crystal with length *L* placed in air (Fig. 1c). In this scenario, the high-index medium acts as an acoustic slab waveguide, guiding sound by total internal refraction as described in [47]. We first aim to perform differentiation, the most fundamental mathematical operation in science and engineering, making use of such simple slab. Assume an incident pressure field with the spatial distribution of $P_i(x)$ is obliquely impinging on the slab as indicated in Fig. 1c. Our goal is to engineer the slab in a way that it carries out differentiation in spatial domain, and provides the derivative of the incoming beam either by reflection or transmission. Considering the transfer function of an ideal differentiator ($G(k_x) = ik_x$), one can argue that, in order to be able to differentiate the input field, the reflection or transmission coefficient of the slab is required to vanish at (at least) one incident angle. Similar to its optical counterpart, the transmission coefficient of this type of acoustic waveguide is non-zero for all incident angles. Nonetheless, its reflectance can, in principle, reduce to zero under certain circumstances. In fact, as outlined in [39], if the length

$L$ satisfies the following condition at some incident angle $\theta_B$, the reflection coefficient will attain a zero value:

$$k_0 n_{eff} \cos(\sin^{-1}(\frac{\sin(\theta_B)}{n_{eff}})) L = \upsilon \pi \quad (1)$$

where $k_0$ is the wavenumber at the operation frequency and $\upsilon$ is an integer. This is nothing but the usual Fabry-Pérot resonance condition, also known as longitudinal transmission resonance in acoustics. What actually happens at this incident angle is that the high-index slab mimics a half-wavelength transmission line, transferring all of the power by matching the characteristic impedances loaded to the input and output of the line. In the following section, we will describe how such a zero in the reflection coefficient of the waveguide can be leveraged for the realization of a spatial differentiation.

Intuitively, one may ask whether such simple waveguide is also able to perform other operators like integration, for example. To answer this question, we should refer to the Green's function of an ideal integrator in the Fourier domain, which is of the form $G(k_x) = \frac{1}{ik_x}$. Depending on whether we choose the transmitted or reflected field as the output, such form of transfer function necessitates the transmission or reflection coefficient of the slab to have a pole at one incident angle. In essence, the reflection coefficient of a slab waveguide cannot have a pole, neither in electromagnetism, nor here in acoustics. Nevertheless, its transmission coefficient can possess poles at certain angles. More specifically, at the incident angles for which the wave-vector and frequency of the incoming beam match those of one of the guided modes of the waveguide, the incident wave strongly excites that mode, creating a pole in the transmission coefficient. This pole in the transmission coefficient of the waveguide can indeed be utilized to carry out spatial integration as we demonstrate in the Results section.

## 3. Results

### 3.1. Spatial differentiation

Consider again the slab waveguide shown in Fig. 1c, and suppose its length is chosen to be *L=9a*. The reflection coefficient of the waveguide is then calculated via full-wave numerical simulation, and is reported in Fig. 2a. As obvious, the reflectance has reduced to zero at $\theta_B \approx$ 30° for which the condition of Eq.1 is fulfilled. We now focus our attention on how the obtained reflection coefficient, calculated for various incident angles, can be mapped onto a transfer function in spatial Fourier domain. As described in [31], one can write the relation between the wave-vector $k_x$ and incident angle $\theta$ as

$$k_x = k_0 sin(sin^{-1}(|sin(\theta)|) - \theta_B) \tag{2}$$

Employing the above equation, one may conveniently map the reflection coefficient onto its corresponding transfer function in wave-vector space. Shown in Fig. 2b is the real part of the transfer function. It is now obvious that the transfer function of the slab is indeed zero at $k_x=0$, which is key for the realization of spatial differentiation. We note that while the transfer function $G(k_x)$ does not exactly match that of an ideal differentiator, it still can be well-estimated with a linear function near the origin as shown in the figure (the red dashed line). As a result, for the incoming fields $P_i(x)$ having sufficiently wide spatial distribution (or equivalently having a small bandwidth in Fourier domain), the reflected field $P_r(x)$ will be very close to the first order derivative of $P_i(x)$. This anticipation can be easily confirmed by considering a spatially wide incident beam having, for example, the Gaussian distribution of in Fig. 2c, and calculating its corresponding reflected field (Fig. 2d). As observed, the reflected field is in perfect agreement with the exact derivative of the incoming field. The profiles of the incident and reflected wave fronts are further analytically calculated and sketched in the inset of Fig. 2e, acknowledging again the proper performance of the differentiator. It should be noted that the amplitude of the reflected field is much lower than that of the incident field. This is actually normal and expected since the transfer functions of differentiators, even in their ideal case, have very low values near the origin.

At this point, it is highly instructive to underline that while the proposed differentiator offers less bandwidth than its MS-based analogues, it is highly compact, a much-sought property allowing its practical usage in integrated architectures. To be specific, the overall length of the slab ($L = \lambda/2$) is about one twenties of that of MS-based differentiators in [40,41]. Furthermore, while the MS-based structures are always accompanied with fabrication complexities due to the coupling responses between metasurfaces, our high-index slab waveguide is quite easy to be implemented in practice [47].

**3.2. Spatial integration**

We next move onto designing an acoustic analog integrator. To this end, all we need is to excite one of the guided modes of the waveguide as explained before in the Methods section. Unfortunately, however, there exists one problem towards achieving such functionality: in principle, it is not possible to excite the guided modes of the slab from air. This is a direct consequence of the phase matching criteria [48], and the fact that the effective refractive index of the slab is higher than its surroundings. In order to get rid of this problem, we use a well-known technique regarded as prism coupling [48,49]. Consider the geometry represented in Fig. 3a, where two additional high-index slabs, acting as the prism couplers, have been put before and after the primary waveguide. Suppose then an obliquely incident wave is travelling inside one of the prisms as schematically shown in the inset of the figure. It is worth noting that the fact that the sound trajectory does not follow the directions of the pipes should not cause any concern as our metamaterial derives its properties from its structure rather than its subwavelength composites. We assume $L_1=4a$, and $L_2=9a$, and calculate the transmission coefficient of the configuration via finite-element numerical simulations (Fig. 3b). The figure reveals the existence of two guided modes whose excitations at the incident angles of $\theta_1 \approx 57°$ and $\theta_2 \approx 68°$ has led to the peaks observed in the transmission. Here, without loss of generality, we focus on the second pole and assume the incident angle of the incoming beam to be $\theta_B = \theta_2$. Similar to the case of the acoustic differentiator, the obtained transmission coefficient can be readily converted to a transfer function in the wave-vector space employing the mapping of Eq. 2. This is actually done in Fig. 3c, indicating the fact that the transfer function of the system can be well-estimated with that of an ideal integrator. We notice, however, that the transfer function is limited to one at the origin, rather than being infinite as it should be for an ideal differentiator. This detrimental property, which stems from the leakage of the excited mode to the prisms, prevents the integrator from retrieving the zero harmonic (or the DC component) of the input field. However, our design still allows the efficient integration of spatially wide AC signals. For instance, suppose a pressure field having the Gaussian-derivative distribution depicted in Fig. 3d as the input of the system. Remarkably, not only is $P_i(x)$ quite wide in the spatial domain but also it contains no DC component. The resulting transmitted field $P_t(x)$ is plotted in Fig. 3e. Comparing $P_t(x)$ with the exact integration of $P_i(x)$ evidences the great functionality of the designed integrator.

### 3.3. Higher order operators

Now that we have successfully performed differentiation and integration, we study the possibility of performing more complex operators. Consider, for example, realization of second order differentiation, an operator which is diversely found in important partial differential equations like wave equation, Schrodinger equation and Euler equation. Our approach to realize this operator is quite straight forward: we cascade two half-wavelength acoustic slab waveguides, each differentiates individually the incoming pressure field one time. The cascading process has been conceptually represented in Fig. 4a: an incident pressure field with the spatial distribution $P_i(x)$ coming in from air strikes the boundary of a half-wavelength slab waveguide. Then, the resulting reflected field $P_{r1}(x)$, which is in fact the spatial derivate of $P_i(x)$, impinges another similar slab waveguide. The final pressure field $P_{r2}(x)$, reflecting back from the second slab will therefore be the second-order derivative of $P_i(x)$, provided that the input signal is sufficiently wide in spatial domain. To examine whether the designed system works well in performing second-order differentiation, let us assume again the input field has the Gaussian distribution of Fig. 2c. The corresponding output field $P_{r2}(x)$ together with the exact second order derivative of the input field are reported in Fig. 4b, confirming the proper functioning of the system. Further insight into the computation process can be obtained by exploring the calculated field profile illustrated in the inset of Fig. 4c.

### 3.4. Edge detection

Finally, we examine the relevance of such acoustic computing devices for detecting the edges of an image. One common and established technique for edge detection, known as zero crossing technique [50], relies on calculating spatial derivative of the image in the direction(s) whose edges are intended to be detected. This can easily be carried out exploiting the acoustic computing setup schematically pictured in Fig. 5a. The setup consists of a loudspeaker, a mask plane with locally engineered transparencies according to the shape of the desire image (shown in Fig. 5b), a half-wavelength acoustic slab waveguide, and a microphone. The underlying working principle of the setup is as follows. The sound generated by the loudspeaker is spatially modulated by the mask plane, creating the appropriate spatial field distribution corresponding to the image. The resulting pressure field then impinges on the high-index acoustic slab and is spatially differentiated when it reflects. The reflected field is then detected by the microphone so as to create the edge-detected image. Fig. 5c reports the output edge-detected image. Notably, since differentiation is carried out along $x$ direction, only the vertical edges have been resolved. The horizontal edges can also be detected employing a similar setup, but with a slab waveguide carrying out differentiation along *y*. The resulting edge-detected image for the latter case is provided in Fig. 5d. One may also consider two-dimensional edge detection by cascading two different differentiators; one differentiates the image along *x* whereas the other does that along *y*. This way, both vertical and horizontal edges of the images can be appropriately detected in one measurement, as observed in the result of Fig. 5e.

## 4. Discussion and Conclusion

In summary, in this article, we proposed to achieve acoustic analog computing making use of high-index slab waveguides. These proposed devices are highly compact and potentially usable in miniaturized (subwavelength) signal processing systems. In particular, our computing devices are superior to their MS-based analogues in that their overall lengths were comparable with the operational wavelength. This is because they directly performed the desired operation in spatial domain, and were needless of bulky Fourier lenses. We demonstrated how a simple, easy-to-fabricate acoustic slab enables realization of spatial differentiators and integrators. We further proved the relevance of such computing devices for detecting the edges of an image. Notice that the edge detection resolution level provided by our wave-based computing systems is inherently restricted by diffraction limit [51-54], while it is not the case when using standard digital image processors. However, our scheme allows one to instantaneously process the entire image at once, which is not achievable with digital data processors. This property can dramatically enhance the throughput of our analog image processor, eventually reducing the cost and energy required for the edge detection process. Furthermore, one may think of improving the design by considering metamaterial slabs supporting spoof acoustic plasmons, which would lead to devices not limited by diffraction and capable of subwavelength edge detection.

These computing devices have the potential to be employed as the building block for fast equation solving as well. For example, consider a very simple first-order differential equation as $\frac{d}{dx}g(x) = f(x)$. To find the unknown function *g(x)*, one should only employ the acoustic setup of Fig. 3a, modulate the incident pressure field in accordance with the known function $f(x)$, and detect the transmitted field. More complex differential or integro-differential equations can also be solved exploiting the cascading method proposed in section 3.3. Such analog equation solvers are not only faster than their digital versions, but also offer low-power consumption operation. Our computing systems can also be envisioned to be utilized in other promising applications including real-time signal processing, wave manipulation and acoustic neural networks.

## Acknowledgement

This work was supported by the Swiss National Science Foundation (SNSF) under Grant No. 172487.

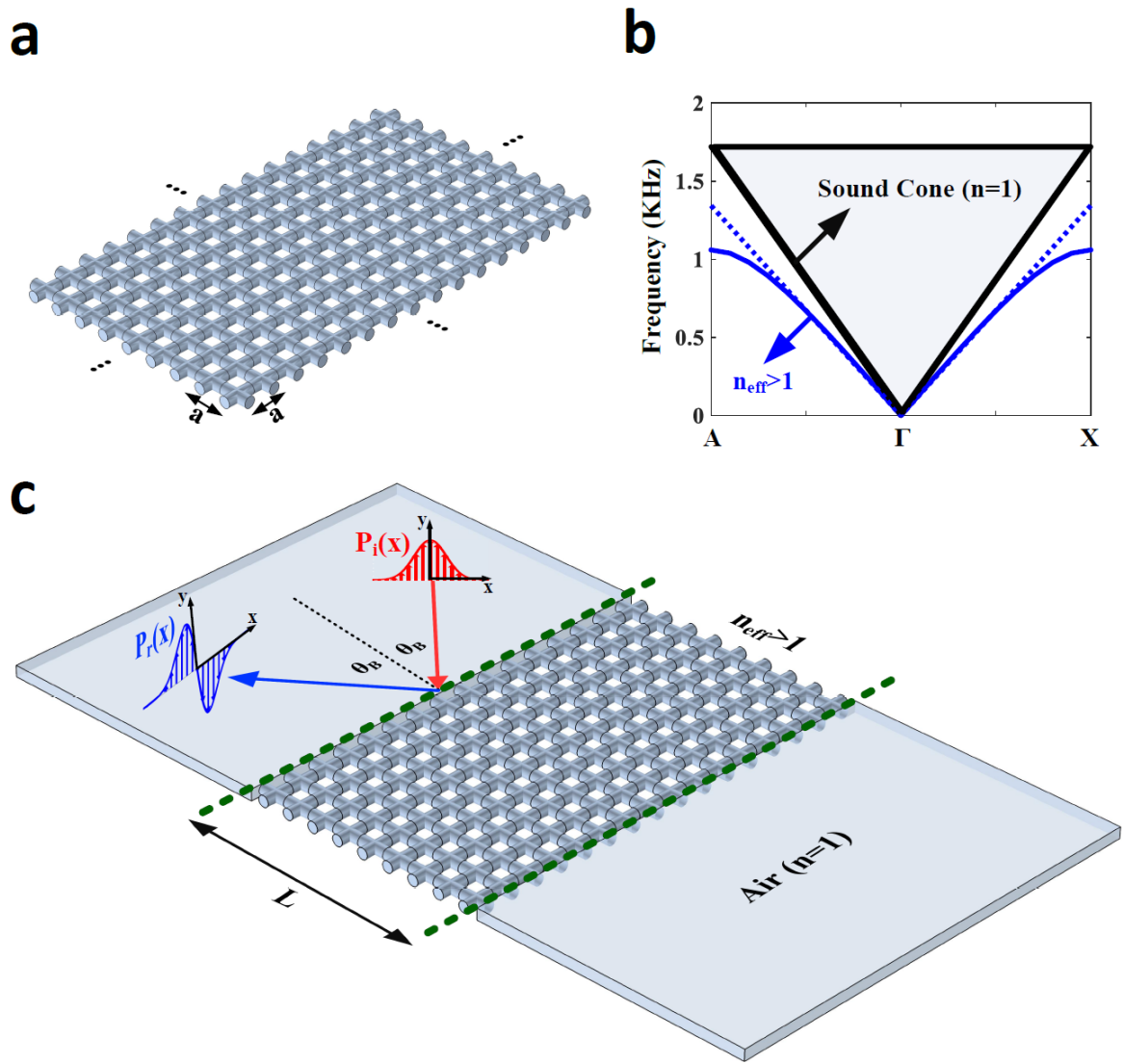

**Fig. 1: Design of an acoustic computing system performing spatial differentiation. a,** Geometry of a periodic acoustic lattice built from air-filled intercrossing pipes. The lattice constant and radii of the pipes are assumed to be *a=5 cm* and *R=1.5 cm*, respectively. **b,** Band structure of the crystal numerically calculated by FEM. The frequency dispersion of the crystal imitates that of a high-index acoustic medium within the frequency range of *0 to 1 KHz*. **c,** An incident pressure field with the arbitrary spatial distribution of $P_i(x)$ impinges a finite piece of the crystal. If the length *L* is chosen to be half of the wavelength of the guiding wave inside the slab, the reflected field will be first-order derivative of the incoming beam.

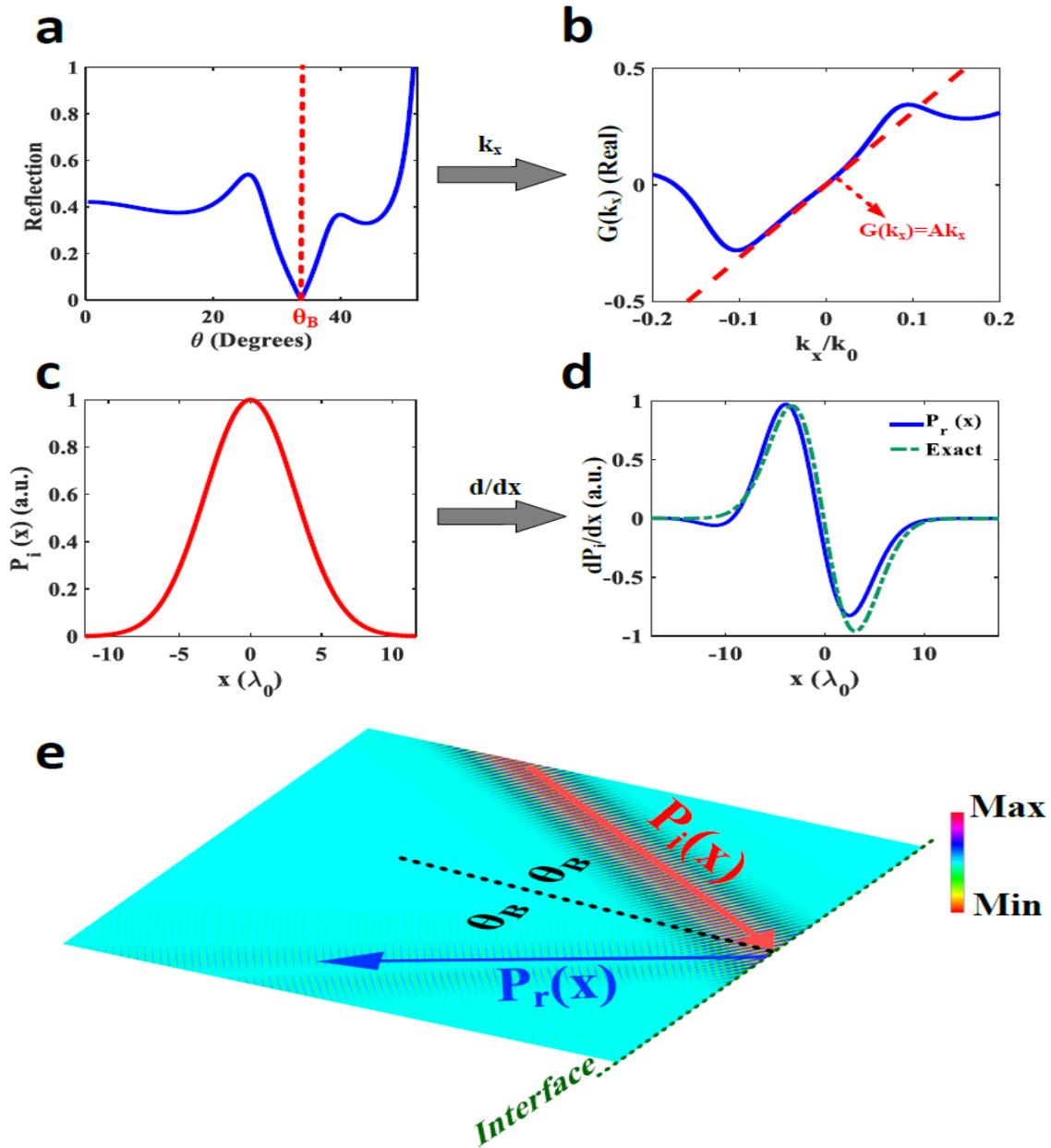

**Fig. 2: Demonstration of proper functioning of the differentiator**, **a,** Variation of reflection coefficient of the waveguide versus the incident angle. The reflection coefficient vanishes at $\theta_B \approx 30°$, forming the underlying working principle of the differentiator. **b,** Transfer function of the differentiator in the wave-vector space obtained using the mapping relation of Eq. 2. **c,** A spatially wide Gaussian-like pressure field is considered as the input signal of the differentiator. **d,** Corresponding reflected field profile: the reflected field follows well the exact derivative of the input signal, affirming the proper functioning of the differentiator. **e,** Field profile of the incident and reflected beams.

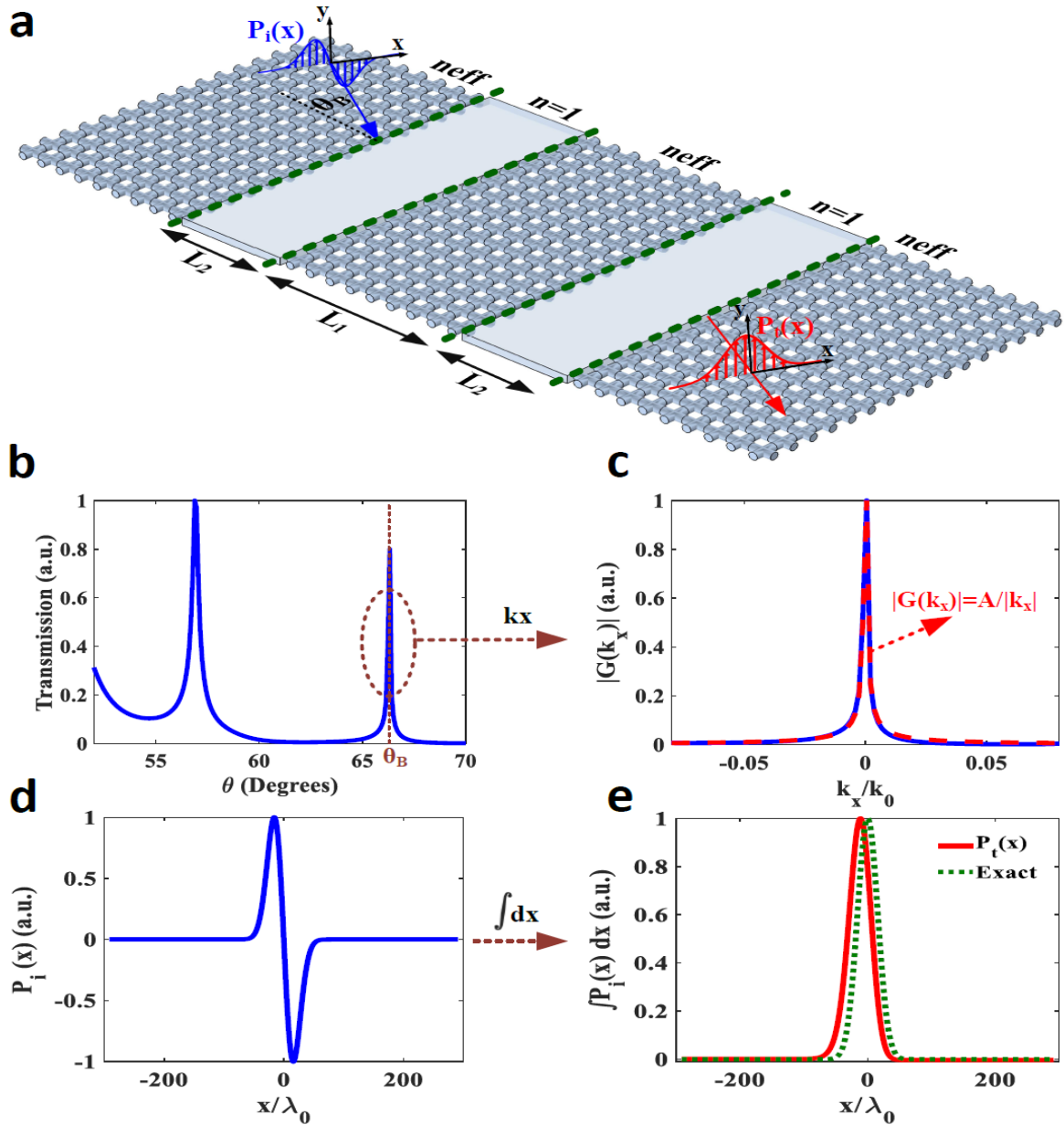

**Fig. 3: Design and demonstration of an acoustic computing system performing integration, a,** Proposed configuration: guided modes of the slab are excited by a pressure field travelling inside another high-index medium acting as a prism coupler. **b,** Variation of transmission coefficient of the system versus the incident angle: the excitation of the guided modes creates some poles in the transmission coefficient, constituting the key underlying working principle of our integrator. **c,** Transfer function of the system. **d,** A spatially wide derivative-Gaussian pressure field is considered as the input signal. **e,** Corresponding transmitted field profile together with the exact integration of the input.

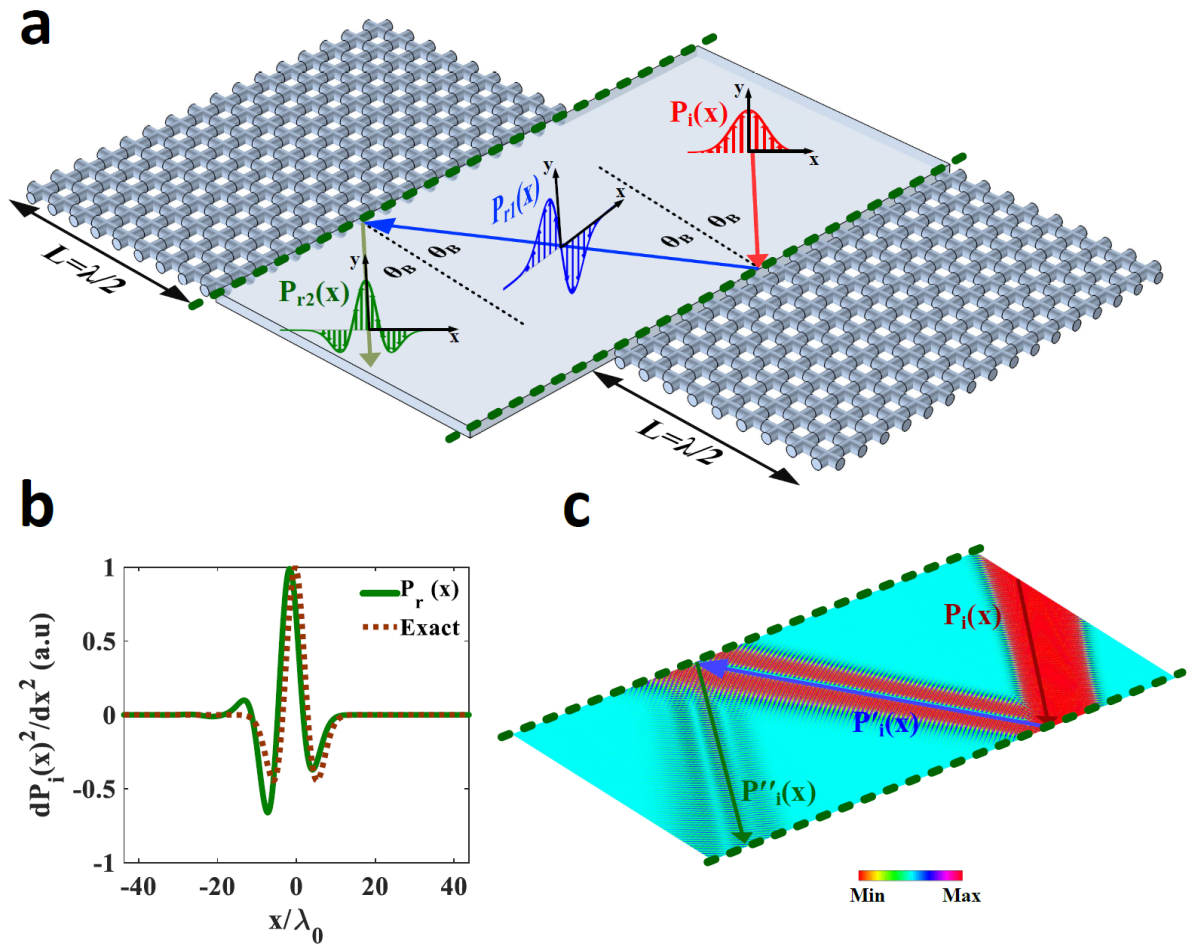

**Fig. 4: Design and Demonstration of an acoustic computing system carrying out second order differentiation, a**, Two half-wavelength slab waveguides are cascaded so as to realize an acoustic second order differentiator. Each waveguide individually differentiates the input field one time. **b,** Output signal of the computing system when the incoming field has the Gaussian-like distribution shown in Fig. 2c. The output field is in perfect agreement with the second order derivative of incident beam. **c,** Field profile of the incident and reflected fields.

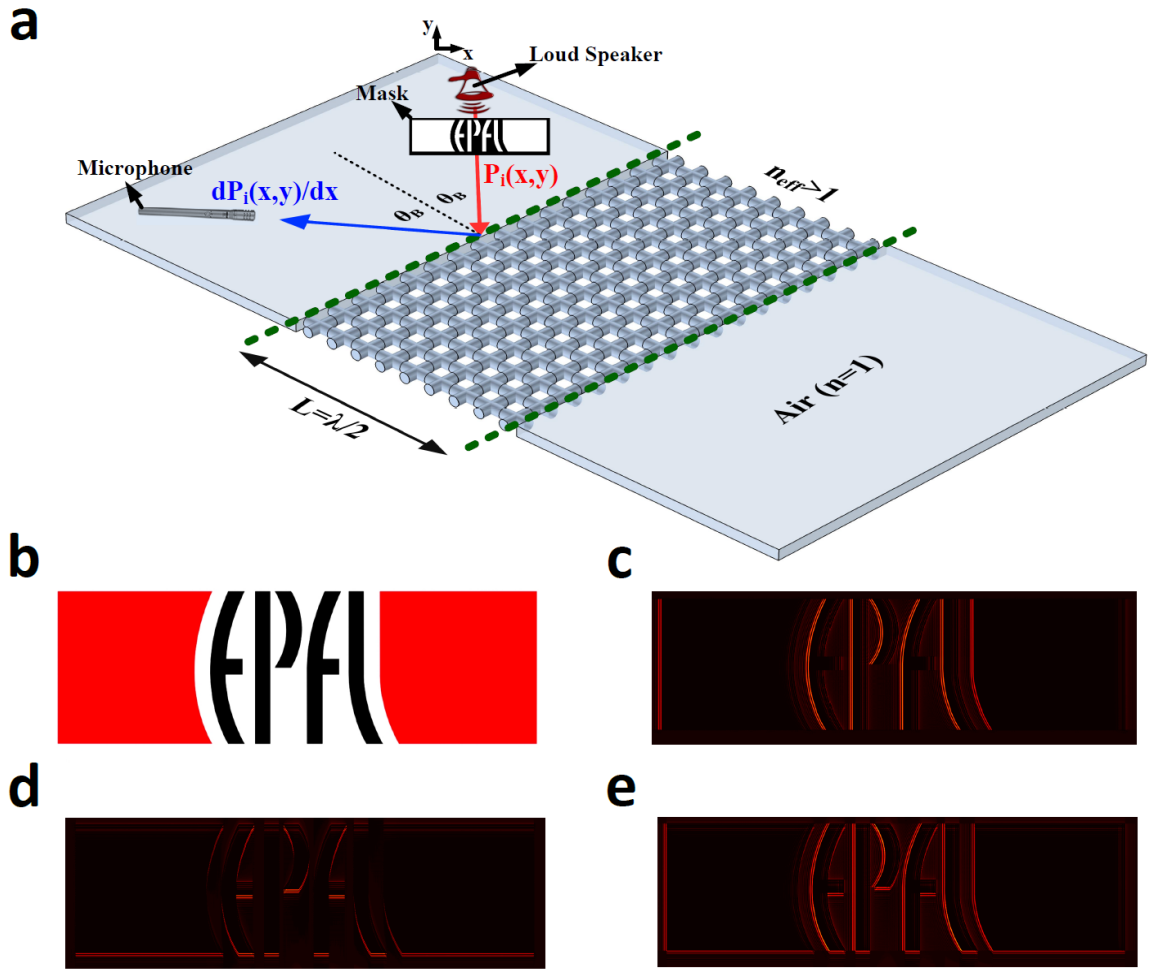

**Fig. 5: Edge Detection of an image utilizing the proposed acoustic computing system, a,** Acoustic computing setup designed for detecting the edges of an image: the setup consists of a loudspeaker as the source, a mask plane with locally engineered transparencies according to the shape of the image whose edges are intended to be detected, a half-wavelength acoustic slab waveguide differentiating the image, and a microphone to resolve edge-detected image, **b,** Photograph of the image whose edges are aimed to be detected, **c,** Output edge-detected image when the slab waveguide differentiates the image in $x$ direction: the vertical edges of the image have been detected. **d,** Same as panel c except that the slab differentiates the image in y direction, detecting its vertical edges. **e,** Resulting image when two cascaded differentiators differentiates the image in both $x$ and $y$ directions, revealing its horizontal and vertical edges simultaneously.

# References

1. Clymer A Ben 1993 The Mechanical Analog Computers of Hannibal Ford and William Newell *IEEE Ann. Hist. Comput.* **15** 19–34
2. Maclennan B J 2007 A Review of Analog Computing *Complex. Syst. Sci. (Springer, 2008)* 1–45C
3. Goodman C 1990 The Digital Revolution: Art in the Computer Age *Art J.* **49** 248–52
4. Smith D R, Pendry J B and Wiltshire M C K 2004 Metamaterials and negative refractive index. *Science* **305** 788–92
5. Alu A and Engheta N 2005 Achieving transparency with plasmonic and metamaterial coatings *Phys. Rev. E - Stat. Nonlinear, Soft Matter Phys.* **72**.
6. Engheta N and Ziolkowski R W 2006 *Metamaterials: Physics and Engineering Explorations*
7. Salandrino A and Engheta N 2006 Far-field subdiffraction optical microscopy using metamaterial crystals: Theory and simulations *Phys. Rev. B - Condens. Matter Mater. Phys.* **74**
8. Chen H and Chan C T 2007 Acoustic cloaking in three dimensions using acoustic metamaterials *Appl. Phys. Lett.* **91** 183518
9. Torrent D and Sánchez-Dehesa J 2007 Acoustic metamaterials for new two-dimensional sonic devices *New J. Phys.* **9**
10. Guenneau S, Movchan A, Patursson G and Ramakrishna S A 2007 Acoustic metamaterials for sound focusing and confinement *New J. Phys.* **9**
11. Fleury R and Alù A 2014 Manipulation of electron flow using near-zero index semiconductor metamaterials *Phys. Rev. B - Condens. Matter Mater. Phys.* **90**
12. Pfeiffer C, Zhang C, Ray V, Guo L J and Grbic A 2014 High performance bianisotropic metasurfaces: Asymmetric transmission of light *Phys. Rev. Lett.* **113**
13. Fleury R, Sounas D L and Alu A 2014 Negative refraction and planar focusing based on parity-time symmetric metasurfaces *Phys. Rev. Lett.* **113**
14. Ross M B, Blaber M G and Schatz G C 2014 Using nanoscale and mesoscale anisotropy to engineer the optical response of three-dimensional plasmonic metamaterials *Nat. Commun.* **5**
15. Sounas D L, Fleury R and Alù A 2015 Unidirectional cloaking based on metasurfaces with balanced loss and gain *Phys. Rev. Appl.* **4**
16. Alù A, Silveirinha M G, Salandrino A and Engheta N 2007 Epsilon-near-zero metamaterials and electromagnetic sources: Tailoring the radiation phase pattern *Phys. Rev. B - Condens. Matter Mater. Phys.* **75**
17. Koutserimpas T T and Fleury R 2018 Zero refractive index in time-Floquet acoustic metamaterials *J. Appl. Phys.* **123**
18. Hwang Y and Davis T J 2016 Optical metasurfaces for subwavelength difference operations *Appl. Phys. Lett.* **109**
19. Cummer S A, Christensen J and Alù A 2016 Controlling sound with acoustic metamaterials *Nat. Rev. Mater.* **1**
20. Yves S, Fleury R, Berthelot T, Fink M, Lemoult F and Lerosey G 2017 Crystalline metamaterials for topological properties at subwavelength scales *Nat. Commun.* **8**
21. Estakhri, N. M., Edwards, B. E., & Engheta, N. (2017, May). Solving integral equations with optical metamaterial-waveguide networks. In *CLEO: QELS_Fundamental Science* (pp. FTh1G-2). Optical Society of America.

22. Abdollahramezani S, Chizari A, Dorche A E, Jamali M V and Salehi J A 2017 Dielectric metasurfaces solve differential and integro-differential equations *Opt. Lett.* **42** 1197
23. Caloz C, Gupta S, Zhang Q and Nikfal B 2013 Analog Signal Processing: A Possible Alternative or Complement to Dominantly Digital Radio Schemes *IEEE Microw. Mag.* **14** 87–103
24. Khorasaninejad M, Chen W T, Devlin R C, Oh J, Zhu A Y and Capasso F 2016 Metalenses at visible wavelengths: Diffraction-limited focusing and subwavelength resolution imaging *Science.* **352** 1190–4
25. Silva A, Monticone F, Castaldi G, Galdi V, Alù A and Engheta N 2014 Performing mathematical operations with metamaterials *Science.* **343** 160–3
26. AbdollahRamezani S, Arik K, Khavasi A and Kavehvash Z 2015 Analog computing using graphene-based metalines *Opt. Lett.* **40** 5239
27. Pors A, Nielsen M G and Bozhevolnyi S I 2015 Analog computing using reflective plasmonic metasurfaces *Nano Lett.* **15** 791–7
28. Solli D R and Jalali B 2015 Analog optical computing *Nat. Photonics* **9** 704–6
29. Golovastikov N V., Bykov D A, Doskolovich L L and Bezus E A 2015 Spatial optical integrator based on phase-shifted Bragg gratings *Opt. Commun.* **338** 457–60
30. Doskolovich L L, Bezus E A, Bykov D A and Soifer V A 2016 Spatial differentiation of Bloch surface wave beams using an on-chip phase-shifted Bragg grating *J. Opt. (United Kingdom)* **18**
31. Youssefi A, Zangeneh-Nejad F, Abdollahramezani S., Khavasi A. 2016. Analog computing by Brewster effect. *Optics letters* **41** 3467-3470
32. Zhu T, Zhou Y, Lou Y, Ye H, Qiu M, Ruan Z and Fan S 2017 Plasmonic computing of spatial differentiation *Nat. Commun.* **8**
33. Doskolovich L L, Bezus E A, Golovastikov N V., Bykov D A and Soifer V A 2017 Planar two-groove optical differentiator in a slab waveguide *Opt. Express* **25** 22328
34. Wu W, Jiang W, Yang J, Gong S, Ma Y 2017 Multilayered analog optical differentiating device: performance analysis on structural parameters *Opt. letters*, **42**, 5270-5273.
35. Fang Y, Lou Y, Ruan Z 2017. On-grating graphene surface plasmons enabling spatial differentiation in the terahertz region. *Opt. letters*, **42**, 3840-3843.
36. Zhang W, Cheng K, Wu C, Wang Y, Li H and Zhang X 2018 Implementing Quantum Search Algorithm with Metamaterials *Adv. Mater.* **30**
37. Guo C., Xiao M., Minkov M., Shi Y., Fan S. 2018 Photonic crystal slab Laplace operator for image differentiation *Optica* **5** 251-256
38. CHEN H, AN D, LI Z and ZHAO X 2017 Performing differential operation with a silver dendritic metasurface at visible wavelengths *Opt. Express* **25** 26417–26
39. Zangeneh-Nejad F, Khavasi A and Rejaei B 2018 Analog optical computing by half-wavelength slabs *Opt. Commun.* **407** 338–43
40. Zuo S Y, Wei Q, Cheng Y and Liu X J 2017 Mathematical operations for acoustic signals based on layered labyrinthine metasurfaces *Appl. Phys. Lett.* **110**
41. Zuo S Y, Tian Y, Wei Q, Cheng Y, Liu X J 2018 Acoustic analog computing based on a reflective metasurface with decoupled modulation of phase and amplitude. *Journal of Applied Physics* **123**, 091704
42. Xie Y, Shen C, Wang W, Li J, Suo D, Popa B I, Jing Y and Cummer S A 2016 Acoustic Holographic Rendering with Two-dimensional Metamaterial-based Passive Phased Array *Sci. Rep.* **6**
43. Tian Y, Wei Q, Cheng Y and Liu X 2017 Acoustic holography based on composite metasurface with decoupled modulation of phase and amplitude *Appl. Phys. Lett.* **110**

44. Liang Z and Li J 2012 Extreme acoustic metamaterial by coiling up space *Phys. Rev. Lett.* **108**
45. Xie Y, Konneker A, Popa B I and Cummer S A 2013 Tapered labyrinthine acoustic metamaterials for broadband impedance matching *Appl. Phys. Lett.* **103**
46. Frenzel T, David Brehm J, Bückmann T, Schittny R, Kadic M and Wegener M 2013 Three-dimensional labyrinthine acoustic metamaterials *Appl. Phys. Lett.* **103**
47. Zangeneh-Nejad F., Fleury R 2018 Acoustic Analogues of High-Index Optical Waveguide Devices *arXiv preprint arXiv:1802.02784.*
48. Maier S A 2007 *Plasmonics: Fundamentals and applications*
49. Zangeneh-Nejad F and Khavasi A 2017 Spatial integration by a dielectric slab and its planar graphene-based counterpart *Opt. Lett.* **42** 1954
50. Haralick R M 1984 Digital Step Edges from Zero Crossing of Second Directional Derivatives *IEEE Trans. Pattern Anal. Mach. Intell.* **PAMI-6** 58–68
51. Zhu J Christensen J, Jung J, Martin-Moreno L, Yin X, Fok L, Zhang X and Garcia-Vidal F J 2011 A holey-structured metamaterial for acoustic deep-subwavelength imaging *Nat. Phys.* **7** 52–5
52. Slaney M, Kak A C and Larsen L E 1984 Limitations of Imaging with First-Order Diffraction Tomography *IEEE Trans. Microw. Theory Tech.* **32** 860–74
53. Zhang S, Yin L and Fang N 2009 Focusing ultrasound with an acoustic metamaterial network *Phys. Rev. Lett.* **102**
54. Lerosey G, De Rosny J, Tourin A and Fink M 2007 Focusing beyond the diffraction limit with far-field time reversal *Science.* **315** 1120–2